\documentclass[11pt]{article}
\usepackage{amsmath}
\usepackage[a4paper, total={6.27in, 9.69in}]{geometry}
\usepackage{graphicx}
\usepackage{hyperref}
\usepackage{textcomp}
\usepackage{gensymb}
\usepackage{placeins}
\usepackage[latin1]{inputenc}
\usepackage{caption}
\usepackage{subcaption}
\usepackage{float}
\usepackage{url}
\usepackage{wrapfig}
\usepackage{lscape}
\usepackage{rotating}
\usepackage{epstopdf}
\usepackage{subcaption}

\begin{document}

\begin{titlepage}
     \begin{center}
     \large
     University of Tripoli\\
     Faculty of Engineering\\
     Department of Electrical and Electronic Engineering\\
     \hfill\break
     \normalsize
     EE491 COMPUTER APPLICATIONS AND DESIGN LAB\\
     \large
     \vspace{8cm}
     \large
     \LARGE
     \textbf{Implementation of DNN Based Data Detector for QPSK Systems\\}
     \Large
     \vspace{10cm}
     \normalsize
     \setlength{\tabcolsep}{20pt}

     \begin{flushleft}
     Written by: Ahmed Mustafa Badi \hfill \\
     ID Number: 22160644 \hfill \\
     Email: eng.a.m.badi@gmail.com\\
     \hfill \break
     Instructor: Dr. Nadia Adam
     \end{flushleft}

     \end{center}
\end{titlepage}

\section*{Abstract}
In this project, the Quadrature Phase Shift Keying (QPSK) digital modulation scheme was implemented using Software Defined Radios (SDRs). For this system, a deep learning based detector was proposed and implemented alongside the conventional method. The implementation was successfully achieved for both the conventional and deep learning based data detection techniques, despite the challenges faced. The results show that the proposed deep learning method is able to outperform the conventional detector. The code of this project is made publicly accessible at \url{https://github.com/ABadi13/QPSK_SDR_DNN_Detector}.

\section{Introduction}
\subsection{Objectives}
To show experimentally and discuss, the implementation of the QPSK digital modulation scheme. Furthermore, to utilize deep learning techniques to design and develop a data detection technique for the QPSK modulation scheme.

\subsection{Background}
As the world's technology becomes ever more sophisticated and powerful, techniques such as Artificial Intelligence (AI) were once thought of as a pipe dream, now AI has been applied to almost every field imaginable. The unique abilities and characteristics of Artificial Neural Networks (ANNs) have made this possible, most notably, their reduced complexities when compared to conventional methods with the same performance. This reduced complexity, not only enables a faster computation time, but also reduced energy consumption and computational requirements. Furthermore, ANNs can provide better awareness of their operational environments, with the ability respond to any imperfections that cannot be accounted for, using conventional methods.

ANNs and Deep Neural Networks (DNNs) have been undergoing substantial advancements in the field of wireless communication, replacing conventional methods in many applications and opening up opportunities in many others. Their reduced complexities allow for cheaper equipment, reduced latency, reduced power consumption, etc. Their greater awareness enables end-to-end learning, allowing them to jointly perform multiple functions, even further simplifying the system and reducing components \cite{web:ai}.

Channel estimation, in particular, has benefited greatly from DNNs. For instance, the authors in \cite{Ye2018} proposed a DNN method for joint channel estimation and symbol detection for OFDM systems. The results of which were promising, with the DNN performing comparably to the conventional channel estimation technique Minimum Mean Square Error (MMSE), and outperforming it in low overhead situations. Additionally, the authors in \cite{Xiang2020} proposed a DNN pair for joint channel estimation and signal detection for the developing spatial modulation scheme. It was concluded that the proposed DNN pair outperforms conventional methods in highly dynamic channel conditions. Instances such as these and many others, show the potential of deep learning in channel estimation.

Cognitive radio is built upon making intelligent decisions to improve the overall utilization of the frequency spectrum. It is a method that allows an unlicensed user to locate and utilize unused licensed spectrum, while also ensuring no interference is caused to the licensed user. One such challenge faced especially in cognitive radio is Automatic Modulation Classification (AMC), in \cite{Meng2018} a Convolution Neural Network (CNN) based AMC approach is proposed. The simulation results showed that the proposed CNN based approach was able to outperform the feature based approach and closely match the much more complex maximum likelihood approach, while maintaining significantly lower computational complexity.

There are many downsides to using DNNs in wireless communication, one important drawback is the DNN's inherent need for large datasets for training. The increased number of layers and neurons need more data to learn the considered system without overfitting. More data is not much of an issue in other fields, although, in wireless communication this is much more difficult. As opposed to regular ANNs, where typically features are manually extracted from the dataset, DNNs typically extract features internally. Meaning that for DNNs, there is a much larger emphasis on the data used.

To combat the large amounts of data, mathematical model-based channels have been utilized to simulate real world environments. Given full Channel State Information (CSI), the model can very effectively estimate the channel. However, the CSI is not always fully available, and some channels can be very dynamic in nature. If the data used to train the DNN is not representative of the real-world channel, then a significant loss in performance is expected.

Software Defined Radios (SDRs), provide the ability to transmit a signal generated and processed in software, or receive a signal to then manipulate it in software. This device practically permits the user freedom in creating any conceivable system, within the realm of the SDRs capability. Real world data can easily be produced using this device, enabling the implementation of ANNs and DNNs directly.

The developments and advancements recently made in SDR technology have been rather rapid in recent years. Several low-cost consumer-focused SDRs have been released such as HackRF One, RTL-SDR, BladeRF, etc. With such success for SDRs, more development and competition is bound to happen in the SDR market, paving the way for technological advancements and accessibility. Academic research utilizing SDR technology has been seen significant success, for instance, the authors in \cite{Zhang2010} have implemented cooperative communication scheme using SDR technology. The results show that the cooperative approach achieves significant performance enhancements. Furthermore, the autors in \cite{Ru2009} presented a digitally enhanced SDR receiver robust to out of bound interference.

More recently, ANNs have been implemented using SDR technology. Instances such as \cite{Jagannath2018} and \cite{Gecgel2019} show the potential of utilizing SDR technology for real world implementation of wireless communication techniques. The authors in \cite{Jagannath2018} designed an ANN based AMC that is able to perform over a wide range of Signal to Noise Ratio (SNR) and then implemented on a SDR test-bed, proving the feasibility of the proposed technique. While in \cite{Gecgel2019} the authors implement two ANNs to jointly determine the presence of a jammer as well as its characteristics. The results in \cite{Gecgel2019} showed that the proposed methods can detect jamming with an $85\%$ accuracy. Despite instances such as this and many others, along with deep learning techniques becoming more relevant than ever in the field of wireless communication, research on the implementation of ANNs leaves much to be desired.

\section{Proposed System Architecture}
This section will outline and discuss the system architectures and algorithms necessary to modulate and demodulate a randomly generated bitstream, using a QPSK digital modulation scheme with a DNN based data detector and implemented using SDRs.

\subsection{QPSK System Architecture}

\begin{figure}
    \centering
    \includegraphics[width=\linewidth]{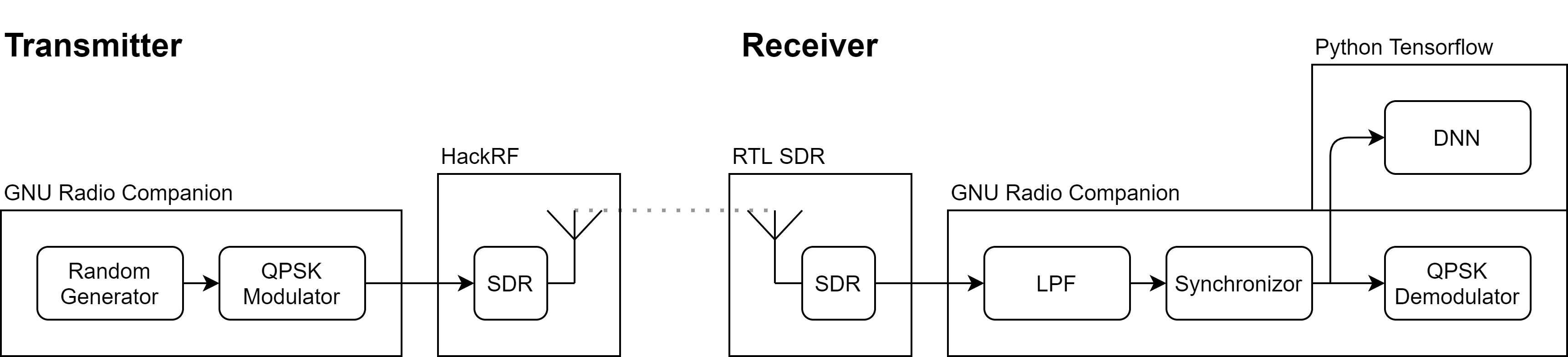}
    \caption{QPSK system utilizing SDR with a DNN based detector}
    \label{fig:sdr_block}
\end{figure}

The QPSK system is illustrated in the form of a block diagram in Fig. \ref{fig:sdr_block}. It is made up of two main sections, the transmitter and the receiver, separated by a wireless channel. Both the transmit and receive SDRs were connected to the same computer, the signal processing, however, will remain entirely independent of each
other, with no feedback loop.

At the transmitter side the GNU Radio Companion software is used to generate a random bit stream, it is then QPSK baseband modulated. The modulated signal is then sent to the \emph{HackRF SDR} to be transmitted over the wireless Radio Frequency (RF) channel.

The implemented channel was a short clear Line Of Site (LOS) channel, about a meter in length. The reason for this choice was due to the SDR's incapability of high power transmission. The sample rate for both SDRs was 2 MHz (samples per second), the lowest common sample rate both SDRs can operate on.

For synchronization and channel estimation, there will be a pilot every frame, this pilot will be omitted when carrying out any performance evaluation. The pilot structure must also remain consistent throughout the entire process.

On the receiver side, the \emph{RTL-SDR} receives the signal sent by the HackRF and then directs it to the GNU Radio Companion software. The received signal is then filtered using an LPF to remove any unnecessary noise. It is then synchronized using a Polyphase Clock Synchronizer and a Costas loop. Once the signal is synchronized, it is stored so that it can be decoded using the conventional method and the proposed DNN. Using the conventional method helps form a baseline for the performance results of the DNN.

\subsection{Algorithm for Creating the Datasets}

\begin{figure}
    \centering
    \includegraphics{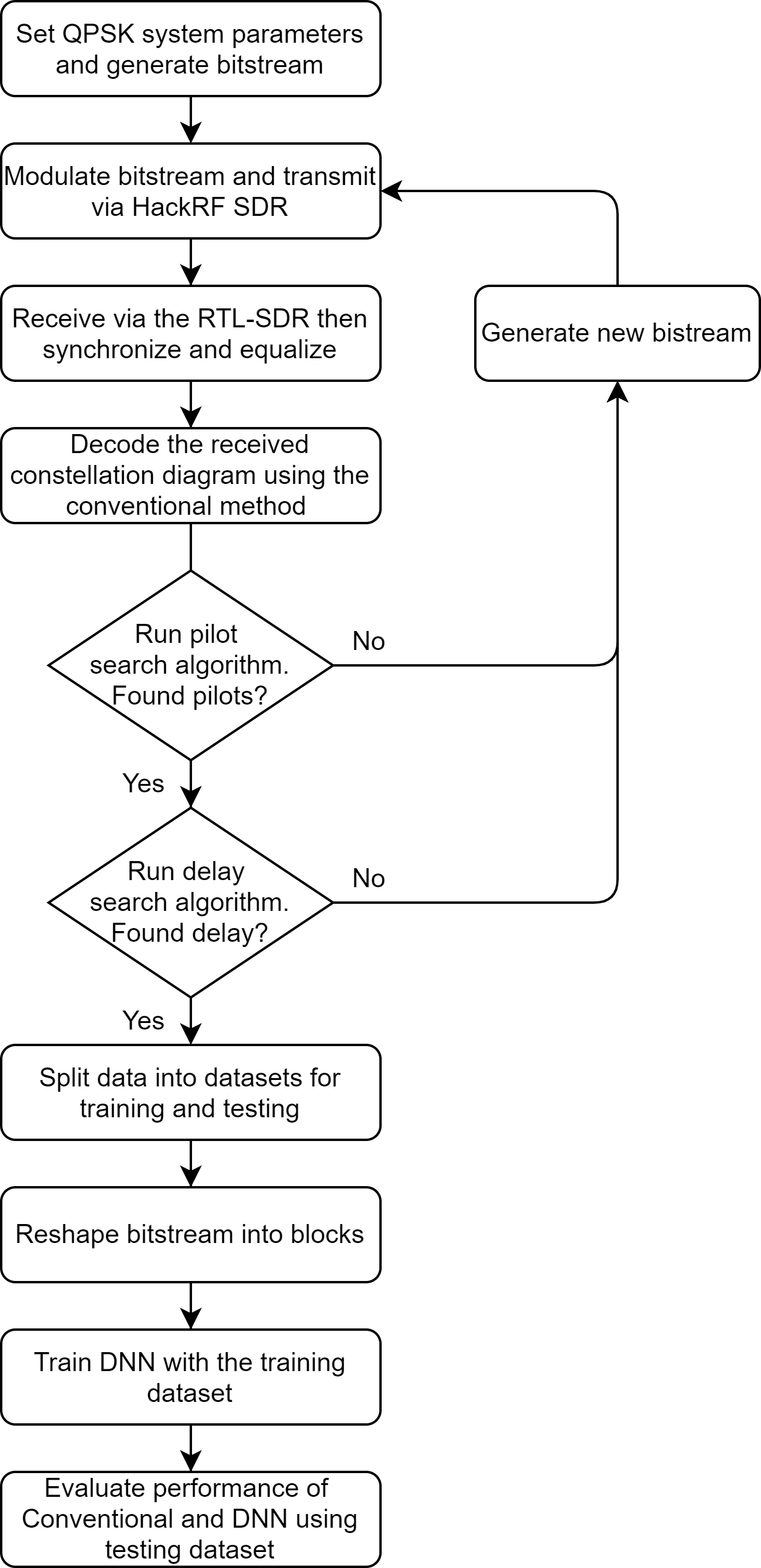}
    \caption{Flowchart outlining the algorithm used for producing the datasets for training and testing.}
    \label{fig:sdr_flow}
\end{figure}

The flowchart shown in Fig. \ref{fig:sdr_flow} illustrates the algorithm this system uses to create datasets and evaluate them using the conventional and the DNN methods. In order to adequately train DNNs with supervised learning techniques, datasets with accurate labels need to be made. These datasets contain the original bitstream, the received bitstream decoded using the conventional method, and the synchronized received signal.

The flowchart starts of as was explained in the previous section, a randomly generated bitstream is modulated, transmitted with the HackRF, received with the RTL-SDR, and then synchronized. 

After this, the synchronized signal is then decoded using conventional methods. Using the decoded bitstream, a pilot search algorithm is run, if no pilots are found then an error has occurred, and the process of transmission is repeated. If the pilots were found, however, another algorithm is run in order to find the delay of the decoded bitstream. If the delay cannot be found then there is an error and the transmission is repeated. If both the phase and the delay can be found, then the training and testing datasets can be created. The training dataset is fed to the DNN to be trained. Then the testing dataset is evaluated using both the conventional and the proposed DNN methods.

It is important to note here that the deployment of a trained DNN is not dependent on the conventional system. It is however, dependent on the conventional system during training to determine only the system delay, in order to accurately label the input samples for supervised learning. It does not use the bitstream decoded using the conventional methods.

\subsection{GNU Radio Companion Flowchart}

\begin{sidewaysfigure}[htbp]
  \centering
  \includegraphics[trim= 0cm 0cm 0cm 0cm, clip, width=\linewidth]{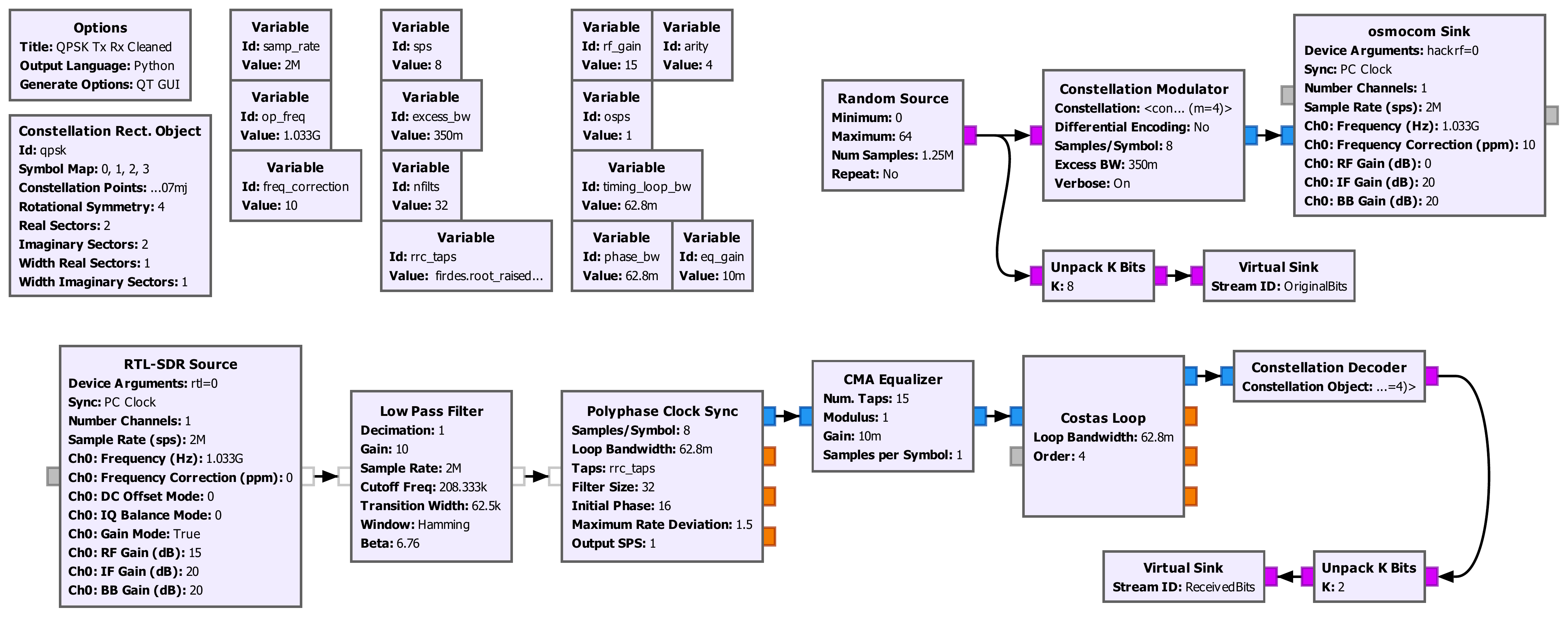}
  \caption{GNU Radio Companion flowchart of the considered QPSK system.}
  \label{fig:gnu_flow}
\end{sidewaysfigure}

The signal processing will be carried out through the open-source software GNU Radio Companion. This software uses blocks that perform basic operations, when connected together these blocks make up a flowgraph, like that shown in Fig. \ref{fig:gnu_flow}. With the flexibility this software allows, a QPSK transmitter and receiver was assembled, each of their corresponding sections are shown in Fig. \ref{fig:gnu_flow}.

Firstly, the random message signal is generated, and for synchronization a pilot is then inserted once every frame. The random bitstream is then QPSK modulated at the baseband using the flexible Constellation Modulator, which modulates according to its constellation object parameter. This block also uses a Root Raised Cosine (RRC) pulse shaping filter to reduce Inter-Symbol Interference (ISI). The modulated signal is then sent to the HackRF SDR which is supported by the OsmoSDR standard. The chosen sample rate was $2$ MHz, with $8$ samples/symbol in order to reduce the signal processing on the computer. The operating frequency was set to $1.033$ GHz for no reason other then being unoccupied.

After the signal passes through the wireless channel, it is received by the RTL-SDR, and then sent back to the computer for processing. An LPF is used in order to remove any unnecessary noise from the received signal. After this, the signal is timing synchronized with the Polyphase Clock Sync. It is equalized using the Constant Modulus Adaptive (CMA) equalizer, which is used for constant amplitude (as the name would suggest) signals, such as the QPSK signal used for this system. Carrier synchronization is achieved using the Costas Loop, the carrier frequency is extracted and the signal is reverted back to the baseband, eliminating any frequency offset experienced by the signal. Finally, after the signal has been synchronized it can be demapped. This signal was then stored and demapped using both, the conventional method utilizing the Constellation Decoder block, and the proposed DNN discussed in the following section. The Unpack blocks are used in order to convert the symbols into bitstreams, so that they can be more easily manipulated when creating the datasets.

\subsection{DNN Architecture}

The proposed DNN system attempts to take the place of the Constellation Decoder. By utilizing the synchronization pilots, the DNN is able to perform channel estimation and potentially improve on the Bit Error Rate (BER) of the conventional method. The proposed DNN will perform the decoding on a frame-by-frame basis. Meaning that the DNN will take the entire frame into consideration when performing the decoding. Its inputs will be all the samples of the synchronized signal in a frame separated into their real and imaginary counterparts. The hidden layers will utilize the \emph{ReLU} activation function. The output layer uses the \emph{Sigmoid} activation function, and attempts to decode the entire frame as bits, excluding any pilots.

Due to the inherent nature of the DNN requiring large amounts of data for training, the network will be deployed in two separate stages, namely, \emph{training} and \emph{testing}. The training stage consists of producing the necessary datasets, this can be achieved by transmitting a priorly known modulated signal, and trained accordingly using supervised learning. After the training stage has been fully fulfilled, the testing stage will be implemented using the produced dataset. In this stage the received modulated signal will be demodulated on a frame-by-frame basis.
\label{ssec:dnn_arc}

\section{Methodology and Challenges}
In this section, the methods used and the challenges encountered during the implementation of the proposed DNN based data detector for the QPSK digital modulation scheme will be outlined and discussed.

\subsection{Deep Learning in Wireless Communications}
Deep learning has seen a substantial amount of research regarding its simulation and how it can be introduced in many applications. Although, real world implementation of machine learning and deep learning of these suggested techniques is not as common as one would assume. Implementing deep learning presents many challenges and potentially allows even further advances in research, and the techniques that can be used to implement it. 

The Python TensorFlow environment was used to train and test the DNN, due to its robustness, speed, and flexibility.

\subsection{The SDR Platform}
Until very recently, implementation using SDRs could only be accomplished with the high-cost Universal Software Radio Peripheral (USRP) devices. Fortunately, low cost SDRs have seen a huge rise in popularity recently, such as the HackRF One, BladeRF, RTL-SDR, etc. that found their way into not only researchers and academics, but also enthusiasts and consumers.

As stated earlier, SDRs allow for the signal processing to be performed through software, opening up many avenues for flexibility in design and configuration. One such option, is the ability to implement machine learning systems, which otherwise would need very specific equipment to implement. However, when using the SDR platform, all that is required is a computer alongside the SDRs to implement the required system.

The SDRs chosen for this system were the \emph{HackRF One} as the transmitter and the \emph{RTL-SDR} as the receiver. The HackRF One device is a half-duplex transceiver, it a is a low-cost open-source hardware project, which would describe its popularity around the SDR platform community. The RTL-SDR on the other hand, is only a receiver, due to its low price, it also has a lot of popularity around the SDR community. Together, they can allow for one way transmission with reasonable pricing.

When implementing custom systems for the SDR platform, there are few software options to choose from, the most popular is the open-source GNU Radio software, due to its wide compatibility for operating systems and SDRs. It is primarily written in Python and C++, and is flexible enough to allow the user to write their own signal processing blocks.

\subsection{Synchronization}
Synchronization is important for any wireless communication system, it helps ensure that the received data actually has meaning and can be utilized, otherwise it may be difficult to make sense of it. Throughout this project, many types of synchronization were encountered and studied such as, timing and carrier synchronization for analog signals, and data synchronization for the decoded digital signal. Despite it being such an important concept, it is difficult to encounter texts that truly address these challenges when performing hardware implementation and how to overcome them. This created a few challenges that were eventually overcome with research and rigorous trial and error.

The timing and carrier synchronization used in this system is the same as that in \cite{web:qpsk_tut}, it was also thoroughly explained. The system in \cite{web:qpsk_tut} in turn became a starting point for the system in this project. Due to the low-cost RTL-SDR, there was a significant frequency offset. To overcome this issue the system was calibrated beforehand, to ensure that the frequency is aligned as best as possible. The timing synchronization, was mainly addressed by \cite{web:qpsk_tut}. However, to fully perform the synchronization and lock onto a phase and the carrier frequency, it takes a short period. During this period, the signal is not synchronized and produces incorrect results when decoded, thus truncation of this period is necessary to ensure that the decoded bitstream was synchronized.

Phase ambiguity, although easy to solve, presented many challenges. Classically, the phase ambiguity is addressed by sending preamble that is known priorly, allowing the bits to be remapped to match the preamble. This works quite well given that the phase lock is consistent, otherwise, if the phase lock may change during operation, pilots (overhead) are required to address this issue, this is called framing synchronization. When the system was first implemented, only a preamble was used, and since the phase lock was inconsistent, especially at low SNR values, this method was not able to properly address this issue. For this reason, pilots were inserted into the bitstream at specific intervals, in order to assist with this synchronization issue.

One of the biggest challenges faced when implementing the SDR system was the delay encountered when transmitting and receiving symbols. Typically, for simple QPSK systems such as that in \cite{web:qpsk_tut}, the delay is expected to be very small and consistent given certain channel conditions, thus, the search for the delay can fairly easily be completed manually. However, for the implemented SDR system, the delay was between 0.1 to 0.3 seconds, and considering that the sampling rate was $2$ MHz, manual search was simply not plausible.

To overcome this issue, the transmitted and received (decoded using the conventional decoder) bitstreams needed to be studied closely, in order to understand what was sent and ensure that the data was truly received without error. A program was then written to iteratively delay the \emph{received} data until it would match the \emph{original} data, this was repeated for several situations and it was concluded that the delay was inconsistent and needs to be found for every dataset created.

In summary, whenever a new dataset is to be created, some of the data at the beginning must be \emph{truncated}, the \emph{phase ambiguity} must be solved, and finally the \emph{delay} must be found. Using all these techniques, datasets for training and testing the DNN were produced.

\section{Implementation Results}

In this section the results of the implementation are shown. The frequency spectrum of the received signal is shown and compared to the transmitted signal. Then the Bit Error Rate (BER) performance of the QPSK system using the conventional and the proposed DNN detection methods are evaluated and compared. Throughout the experiment, the channel was kept stationary at a length of approximately $1$ m, with a clear LOS between the transmitter and the receiver. The sample rate was set to $2$ MHz and the samples per symbol was set to $8$, in order to reduce computation requirements of the system. A frame size of $4$ symbols was also considered. The proposed DNN uses the framework discussed in section \ref{ssec:dnn_arc}, had five layers of sizes $8$, $100$, $50$, $20$, and $6$, respectively.

\subsection{The Transmitted and The Received Signal}

\begin{figure}
    \begin{subfigure}{\textwidth}
        \includegraphics[width=\textwidth]{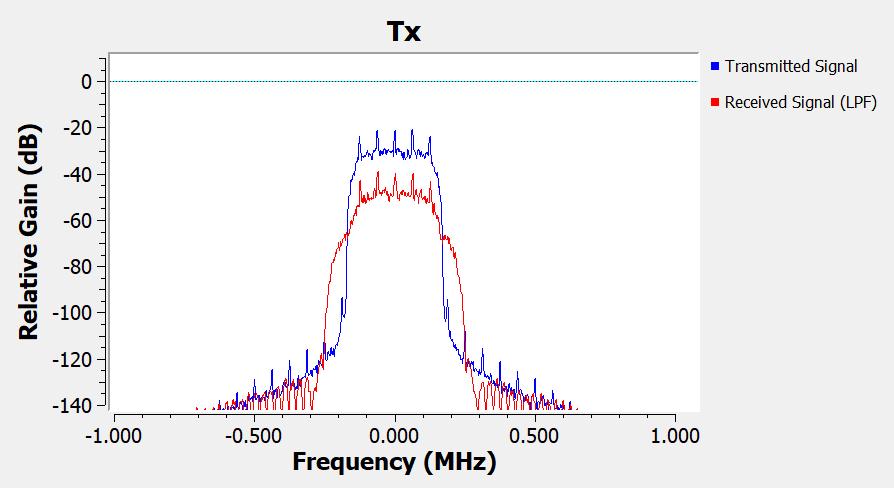}
        \caption{All frequency components received by the SDR.}
        \label{fig:txrx}
    \end{subfigure}
    \begin{subfigure}{\textwidth}
        \includegraphics[width=\textwidth]{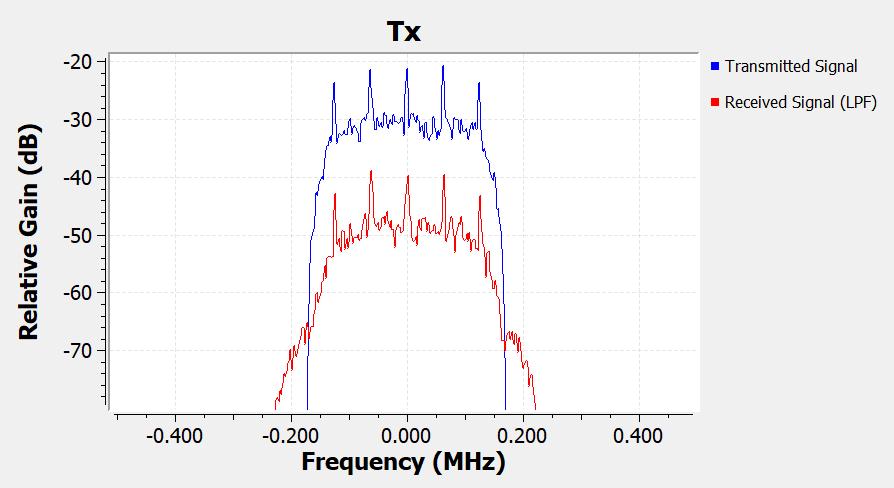}
        \caption{Zoomed in for a better view of the modulated signal}
        \label{fig:txrx_zoom}
    \end{subfigure}
    \caption{Frequency spectrum of the transmitted signal and the received signal after the LPF.}
\end{figure}

Figures \ref{fig:txrx} and \ref{fig:txrx_zoom} show the frequency spectrum of the transmitted signal along with the received signal after being low pass filtered. It is recognized from the transmitted signal that there is a steep dropoff in the magnitude of the frequency components. This is owed to the RRC pulse shaping filter, which in turn is used for timing synchronization for optimal sampling, reducing ISI. The bandwidth of this signal can be increased by increasing the Excess BW parameter of the Constellation Modulator block. Ideally, there would be no frequency components after the bandwidth of the signal, however, due to it being processed digitally, the floating point precision comes into play and introduces high frequency components that are mostly negligible.

These figures show how well the signal was received with very little distortion, owing to the short LOS channel. Also only a slight frequency offset is noticed, due to the SDRs being calibrated beforehand. From the figures, five impulses are also observed at both the transmitted and received signals. These impulses are a result of the pilot symbol being sent so frequently, increasing the power received from the pilots relative to the rest of the symbols.

\begin{figure}
    \begin{subfigure}{\textwidth}
        \includegraphics{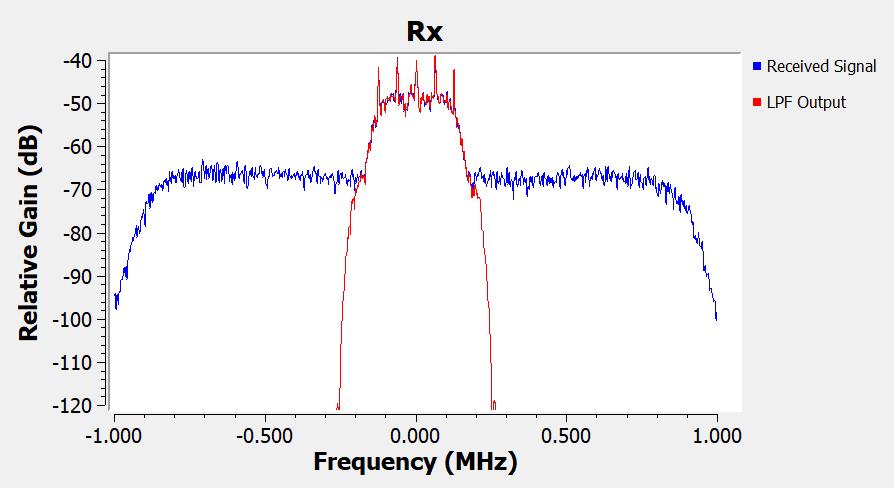}
        \caption{All frequency components received by the SDR.}
        \label{fig:rx}
    \end{subfigure}
    \begin{subfigure}{\textwidth}
        \includegraphics{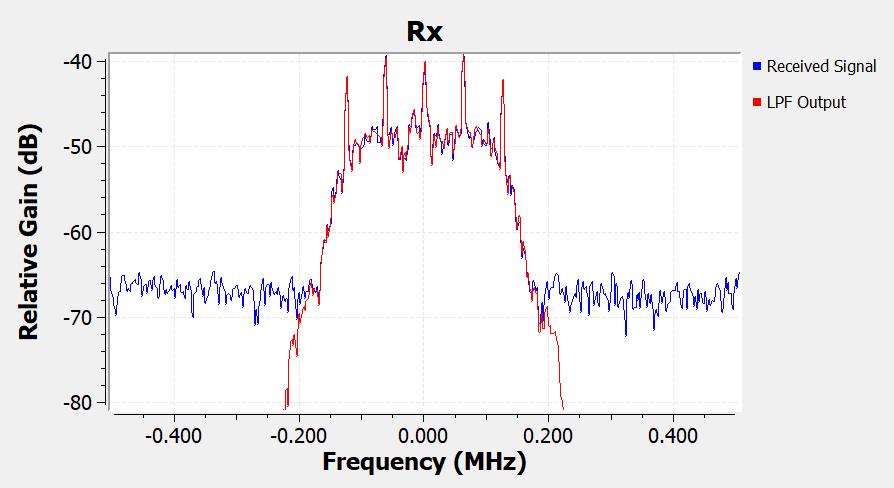}
        \caption{Zoomed in for a better view of the modulated signal.}
        \label{fig:rx_zoom}
    \end{subfigure}
    \caption{Frequency spectrum of the received signal before and after being low pass filtered.}
\end{figure}

Shown in Fig. \ref{fig:rx} is the LPF output compared to the received signal. Before around 200 kHz the two signals very closely resemble each other, above this frequency however, heavy attenuation from the LPF eliminates high frequency components, reducing noise and allowing for more accurate demodulation.

\subsection{Constellation Diagram}
\begin{figure}
    \centering
    \includegraphics[width=0.8\textwidth]{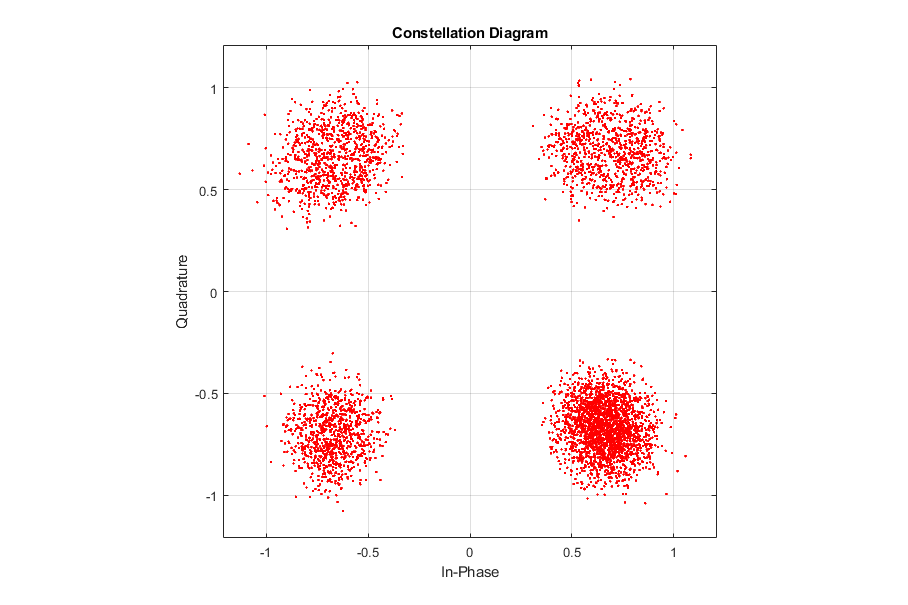}
    \caption{Constellation diagram of the received signal.}
    \label{fig:const}
\end{figure}

The constellation diagram at an SNR of approximately 19 dB is shown in Fig. \ref{fig:const}, it shows the in-phase and quadrature components of the synchronized signal. From this constellation diagram, the clusters can be viewed clearly and separated very easily. With careful inspection, it also shows a nonuniformity in the distribution of received symbol in each quarter. In other words, one of the received symbols is received far more frequently than the others, this is quite clearly due to the pilot symbol, which takes up to a quarter of the frame time for the considered system.

\subsection{BER Performance}

\begin{figure}[!htbp]
    \centering
    \includegraphics[width=0.8\textwidth]{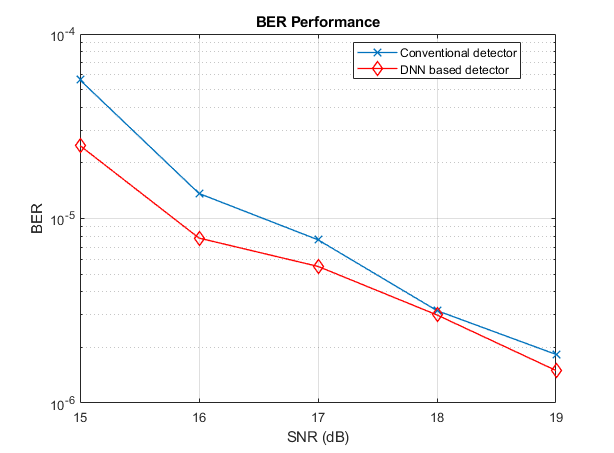}
    \caption{BER performance of the proposed DNN data detector.}
    \label{fig:ber}
\end{figure}       

The BER performance of the conventional and the proposed DNN approaches are shown in Fig. \ref{fig:ber}. The BER performance is measured against variation in the SNR value of the signal. The SNR is measured by finding the difference between the noise floor and the average power of the received signal, which could then be adjusted by altering the RF gain value of the RTL-SDR.
\begin{equation}
    SNR = P_{R}+G_{R}-N
    \label{eq:snr}
\end{equation}
where $P_{R}$ is the received average power with no receiver gain, $G_R$ is the gain of the receiver, and $N$ is the noise floor. From Fig. \ref{fig:rx_zoom} the noise floor was measured to be approximately $N=-68$ dB, and the average power of the received signal is approximately $P_R+G_R=-49$ dB with the gain of the RTL-SDR set to $G_R=15$ dB. Therefore, assuming that both the noise floor and the received power without receiver gain remain constant, the SNR can be approximately measured directly from the gain of the receiver as such
\begin{equation}
    SNR = 4 \text{ dB} + G_R
\end{equation}

It is observed from this figure that the proposed DNN clearly outperforms the conventional approach. Proving the potential of DNNs to perform well under real world channels. The figure also shows that the smooth graphs seen in simulation may not be as easy to maintain in real world channels, further proving the need for implementation studies.

\section{Conclusions and Future Work}
In this project, a DNN based data detection method for the QPSK digital modulation scheme was designed and implemented using the HackRF One and the RTL-SDR. The implementation results show that the signal was received well, with distinctive clusters for each symbol. The BER performance was also evaluated, it showed that the DNN based detector outperforms the conventional method for the considered system. It was also recognized that implementation presents numerous challenges not encountered when performing simulations, fortunately these challenges were overcome and discussed in this report.

Future work will expand this methodology to different digital modulation schemes, such as Amplitude Shift Keying (ASK) and Frequency Shift Keying (FSK). Spatial modulation can also be implemented, however, doing so would require multiple transmit SDRs. Furthermore, cognitive radio applications present numerous avenues to explore, such as modulation classification, reconfigurable smart transceivers, spectrum sensing, etc. The flexibility of SDRs and the GNU Radio software allow for endless applications.

\section{Acknowledgement}
This work was partially funded by the Libyan Foundation for Projects Development.

\end{document}